\documentclass[twocolumn,seceq]{jpsj3}
\usepackage{graphicx}

\def\runauthor{H. {\sc Ueda}, A. {\sc Gendiar}, and T. {\sc Nishino}}

\title{A Classical Background for the Wave Function Prediction in 
the Infinite System Density Matrix Renormalization Group Method}

\author
{Hiroshi {\sc Ueda}$^{1)}$, Andrej {\sc Gendiar}$^{2)}$, and 
Tomotoshi {\sc Nishino}$^{3)}$}
\inst
{$^1$Department of Material Engineering Science, Graduate School of 
Engineering Science, Osaka University, Osaka 560-8531, Japan\\
$^2$Institute of Electrical Engineering, Slovak Academy of Sciences,
D\'ubravsk\'a cesta 9, SK-841 04, Bratislava, Slovakia\\
$^3$Department of Physics, Graduate School of Science, 
Kobe University, Kobe 657-8501, Japan
}

\abst
{We report a physical background of the wave function prediction
in the infinite system density matrix renormalization group (DMRG) method, 
from the view point of two-dimensional vertex model, a typical lattice 
model in statistical mechanics.
Singular value decomposition applied to rectangular corner transfer 
matrices naturally draws matrix product representation for the maximal 
eigenvector of the row-to-row transfer matrix. The wave function prediction 
can be expressed as the insertion of an approximate half-column transfer 
matrix. This insertion process is in accordance with the scheme proposed 
by McCulloch recently.}

\kword{DMRG, PWFRG, CTMRG, Renormalization}

\begin{document}
\sloppy
\maketitle

\section{Introduction}

The density matrix renormalization group (DMRG) method is one
of the efficient numerical method, which has been applied extensively 
to one-dimensional (1D)
quantum systems and two-dimensional (2D) classical systems.~\cite{DMRG,
DMRG2,int,DMRG3} 
The method is variational in the sense that it assumes a trial state,
the matrix product state (MPS), which is written as a product of local 
tensors.~\cite{AKLT,Fannes1,Fannes2,Fannes3,Klumper1,Klumper2,
Ostlund,Ostlund2,Ostlund3,Takasaki,Sierra} Orthogonality of each matrix 
ensures the numerical stability. 

One of the bottleneck in the computation of the DMRG method is the 
diagonalization of super block Hamiltonian. The construction of a good
initial vector for this diagonalization is very important. For the finite-system
DMRG method, the so-called wave function renormalization scheme
provides the answer.~\cite{Acce,Acce2} For the infinite-system DMRG method, 
Baxter's method of corner transfer matrix (CTM),~\cite{Baxter1,
Baxter2,Baxter3} which can be reinterpreted from the view point of the 
DMRG method,~\cite{CTMRG,CTMRG2} essentially solves 
the problem of initial vector. Based on Baxter's CTM method, the product
wave function renormalization group (PWFRG) method was 
proposed,~\cite{PWFRG,PWFRG2,PWFRG3} and has been applied to the study of 1D 
spin chains.~\cite{Acts,Acts2,Acts3,Acts3-2,Acts3-3,Acts3-4,Hieida,
Hagiwara,Okunishi,Okunishi-2,Okunishi-3,Okunishi-4,Okunishi-5,
Narumi,Yoshikawa}

Recently McCulloch proposed a way of precise wave function prediction,
which works better than the PWFRG method especially when the system 
size is small compared with the correlation length.~\cite{McCulloch} 
In this paper we present a physical background for McCulloch's scheme
from the view point of 2D vertex model, one of the typical lattice model
in statistical mechanics.~\cite{Baxter3} Although we employ classical
lattice model, most of the obtained results can be applicable for 1D quantum
systems through the quantum-classical correspondence.

Structure of the paper is as follows. In the next section we 
explain the symmetric vertex model, and express the maximal
eigenstate of the row-to-row transfer matrix by use of CTMs.
In \S 3 we consider the area extension of CTMs, introducing
an approximate half-column transfer matrix. We show the
connection between MPS and CTM formulation in \S4, where
the system size extension scheme by McCulloch is obtained
naturally. We summarize the obtained result in the last section, and
discuss the remaining problem on the MPS obtained by
the finite-system DMRG method.

\section{Eigenstate of row-to-row Transfer Matrix Approximated
by Corner Transfer Matrices}

\begin{figure}
\centerline{\includegraphics[width=6.5cm,clip]{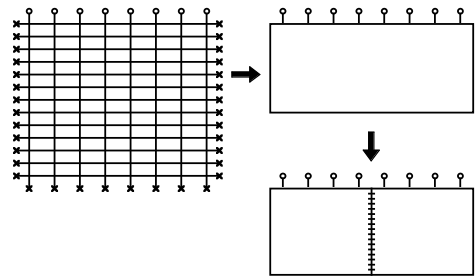}}
\caption{ A finite size vertex model of width $N = 8$. Cross marks 
show boundary spins and open circles show row spins, which are
the variables of $\Psi_8^{~}$ in Eq.~(2.1). The system consists of
its left-half and the right half, where the vertical stitch corresponds 
to the half-column spin $\sigma_1^{~}, \sigma_2^{~}, \sigma_3^{~}, \ldots$
between them.}
\label{f1}
\end{figure}

Throughout this article we consider a square-lattice symmetric vertex 
model,~\cite{Baxter3} as an example of 2D classical lattice models. There is a $d$-state 
spin variable on each bond, which connects neighboring lattice points. 
Four spins around a lattice point 
determine the local Boltzmann weight $W$, which is called as the vertex 
weight. We assume that the vertex weight is position independent, 
and therefore the system is uniform. We also assume that each vertex 
weight is invariant under exchange of left and right spin variables, and
those of up and down spin variables. In other words, we consider the
symmetric vertex model in order to simplify the following formulation.

As shown on the left side of Fig.~1, we treat a finite size system that has 
a rectangular shape. This system corresponds to the stack of 
row-to-row transfer matrices $T_N^{~}$, whose width is $N$, multiplied by an 
initial vector $V_N^{~}$. We choose $V_N^{~}$ so that it corresponds to
the boundary condition at the bottom of the system, where there is a 
row of boundary spins shown by the cross marks. Those cross marks aligned 
vertically also represent boundary spins, that are located 
at the both ends of $T_N^{~}$. 
The row of open circles represents spins on top of the rectangular system. 
We consider a $d^N_{~}$-dimensional vector
\begin{equation}
\Psi_N^{~} = T_N^{~} T_N^{~} T_N^{~} \ldots T_N^{~} V_N^{~} \, ,
\end{equation}
where the number of the row-to-row transfer matrix $T_N^{~}$ is sufficiently 
large. Under this assumption we can expect that $\Psi_N^{~}$ is a good 
approximation of the maximal eigenvector of $T_N^{~}$ if $V_N^{~}$ is
not orthogonal to that.

For a while let us consider the case $N = 8$; generalization to arbitrary 
$N$ is straightforward. 
We label the top spins as $q_1^{~}, \, q_2^{~}, \, q_3^{~}, \, q_4^{~}, \, 
p_4^{~}, \, p_3^{~}, \, p_2^{~}$, and $p_1^{~}$ from left to right. 
The vector elements of $\Psi_8^{~}$ are then written as 
$\Psi_8^{~}( q_1^{~}, \, q_2^{~}, \, q_3^{~}, \, q_4^{~}, \, 
p_4^{~}, \, p_3^{~}, \, p_2^{~}, \, p_1^{~} )$. Since we have assumed
the left-right symmetry for the vertex weight, it is convenient to divide 
the row-spin into the
left half $q_1^{~}, \, q_2^{~}, \, q_3^{~}, \, q_4^{~}$, where we have 
counted them from left to right, and the right half $p_1^{~}, \, p_2^{~}, 
\, p_3^{~}, \, p_4^{~}$, where we have counted them from right to left.
(See right bottom of Fig.~1.) According to this division, we can 
interpret $\Psi_8^{~}$ as a $d^4_{~}$-dimensional real symmetric matrix, 
whose elements can be expressed as 
$\Psi_8^{~}( q_1^{~} \, q_2^{~} \, q_3^{~} \, q_4^{~} | 
p_1^{~} \, p_2^{~} \, p_3^{~} \, p_4^{~} )$. We have used the vertical 
bar ``$|$'' to separate the left and the right indices, and
dropped the commas between the spin variables for the book keeping.
If necessary, we further abbreviate the matrix notation as $\Psi_8^{~}( q | p )$. 

We express the left half of the rectangular system by use of the 
CTM, whose elements are written as 
$C_4^{~}( q_1^{~} \, q_2^{~} \, q_3^{~} \, q_4^{~} |
\sigma_1^{~} \, \sigma_2^{~} \, \sigma_3^{~} \, \ldots )$, where 
$\sigma_1^{~} \,  \sigma_2^{~} \, \sigma_3^{~} \, \ldots$ represent 
the half-column spins at the center of the system. In the same manner
we can express the right half by the transpose of $C_4^{~}$, i.e., $C_4^{~T}$.
Joining process of these halves by stitching $C_4^{~}$ and $C_4^{~T}$ via the 
contraction of the half-column spins can be expressed simply by the 
product of matrices $\Psi_8^{~} = C_4^{~} \, C_4^{~T}$. More precisely,
there is a relation
\begin{equation}
\Psi_8^{~}( q | p ) = \sum_{\sigma}^{~} 
C_4^{~}( q | \sigma ) \, 
C_4^{~}( p | \sigma ) \, ,
\end{equation}
where we have used the abbreviations  $q = q_1^{~} \, q_2^{~} \, q_3^{~} \, q_4^{~}$, 
$\, p = p_1^{~} \, p_2^{~} \, p_3^{~} \, p_4^{~}$, and
$\sigma = \sigma_1^{~} \sigma_2^{~} \sigma_3^{~} \ldots \,\,\, $.
Since we have assumed that the number of $T_8^{~}$ in Eq.~(2.1) is sufficiently 
large, the same for the number of column-spin $\sigma$. Although we 
treat $\sigma$, we do not think of them as spins directly treated in numerical 
calculations, unlike $q$ and $p$. 

\begin{figure}
\centerline{\includegraphics[width=5.5cm,clip]{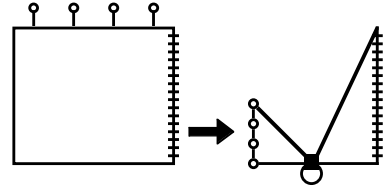}}
\caption{ The singular value decomposition applied to $C_4^{~}$. 
The black square and circle corresponds to the block spin
$\xi$ and the singular values $\Omega_4^{~}( \xi )$, respectively. 
The left and the right triangles represent $A_4^{~}$ and $U_4^{~}$, respectively.}
\label{f2}
\end{figure}

One of the fundamental mathematical tool in the DMRG method is the singular value
decomposition (SVD).~\cite{DMRG,DMRG2} Let us apply it to the CTM
\begin{equation}
C_4^{~}( q | p ) = \sum_{\xi}^{~} A_4^{~}( q | \xi ) 
\, \Omega_4^{~}( \xi ) \,
U_4^{~}( \sigma | \xi ) \, ,
\end{equation}
where $\xi$ is a $d^4_{~}$-state block-spin (or an auxiliary) variable, and 
$\Omega_4^{~}( \xi )$ represents the singular values.
The matrix $A_4^{~}$ is $d^4_{~}$-dimensional, and it
satisfies the orthogonal relations
\begin{eqnarray}
\sum_\xi^{~} A_4^{~}( q' | \xi ) \, A_4^{~}( q | \xi ) 
\!\!\! &=& \!\!\! \delta( q' | q ) \, ,
\nonumber\\
\sum_q^{~} A_4^{~}( q | \xi' ) \, A_4^{~}( q | \xi ) 
\!\!\! &=& \!\!\! \delta( \xi' | \xi ) \, ,
\end{eqnarray}
where $\delta( \xi' | \xi )$ is Kronecker's delta, and
where $\delta( q' | q )$ is defined as 
\begin{equation}
\delta( q' | q ) = \prod_{i=1}^{4} \delta( q'_i | q_i^{~} ) \, .
\end{equation}
The above orthogonal relation can be written shortly as 
$A_4^{~} A_4^{~T} = A_4^{~T} A_4^{~} = I_4^{~}$. 
Column vectors of the rectangular matrix $U_4^{~}$ are
also orthogonal with each other,
\begin{equation}
\sum_{\sigma}^{~} U_4^{~}( \sigma | \xi' ) \, U_4^{~}( \sigma | \xi ) 
= \delta( \xi' | \xi ) \, ,
\end{equation}
but the row vectors are not
\begin{equation}
\sum_\xi^{~} U_4^{~}( \sigma' | \xi ) \, U_4^{~}( \sigma | \xi ) 
\neq \delta( \sigma' | \sigma ) \, .
\end{equation}
This is because the degree of freedom of $\sigma$ is far larger than
that of $q$ or $\xi$.
Figure 2 is the pictorial representation of SVD applied to $C_4^{~}$.

We often regard the singular values $\Omega_4^{~}$ as the diagonal matrix
$\Omega_4^{~}( \xi' | \xi ) = \Omega_4^{~}( \xi ) \, 
\delta( \xi' | \xi )$, and write Eq.~(2.3) 
shortly as $C_4^{~} = A_4^{~} \, \Omega_4^{~} \, U_4^{~T}$.
For the latter convenience, let us introduce the generalized inverse
of the CTM
\begin{equation}
C_4^{-1} =U_4^{~} \, \Omega_4^{-1}  A_4^{~T} \, ,
\end{equation}
which satisfies the relation
\begin{eqnarray}
C_4^{~} \, C_4^{-1} \!\!\! &=& \!\!\! 
A_4^{~} \, \Omega_4^{~} \, U_4^{~T}
U_4^{~} \, \Omega_4^{-1}  A_4^{~T} \nonumber\\
\!\!\! &=& \!\!\!  A_4^{~} \, A_4^{~T} = I_4^{~} \, .
\end{eqnarray}
It should be noted that $C_4^{-1} C_4^{~}$ is
a projection operator
\begin{equation}
U_4^{~} \, \Omega_4^{-1} A_4^{~T} A_4^{~} \, \Omega_4^{~} \, U_4^{~T}
 = U_4^{~} \, U_4^{~T}
\end{equation}
in the left hand side of Eq.~(2.7), where $( C_4^{-1} C_4^{~} )^2_{~} = 
C_4^{-1} C_4^{~}$ holds.

In the context of the DMRG method, small singular values are neglected when it is
impossible to store matrix elements during the numerical calculation. 
This truncation is a kind of decimation in the renormalization group (RG) 
theory. Under the truncation, the matrices $A_4^{~}$ work as the 
RG transformation that controls numerical precision. 
In the next section we do not truncate singular values, in order to avoid 
complications in notations, and the introduction of truncation is straightforward.

\section{Half Column Transfer Matrix and Matrix Product State}

\begin{figure}
\centerline{\includegraphics[width=5.5cm,clip]{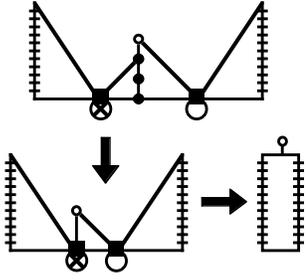}}
\caption{ The pictorial representation of $P_4^{~} = C_3^{-1} \! \cdot  C_4^{~}$
in Eq.~(3.1). The circle with the cross mark shows $\Omega_3^{-1}$. 
Since $P_4^{~}$ has a function of extending the area of CTM, it can be
regarded as an approximation for the half-column transfer matrix.}
\label{f3}
\end{figure}

We introduce a new notation between matrices, the dot product, which contract 
variables according to Einstein rule. As an example, let us consider 
\begin{equation}
P_4^{~} = C_3^{-1} \! \cdot  C_4^{~} \, ,
\end{equation}
where $q_1^{~}$, $q_2^{~}$, and $q_3^{~}$ are contracted but $q_4^{~}$ is not, 
since the first three spins are shared by $C_3^{-1}$ and
$C_4^{~}$. Figure 3 shows this rule graphically.
Substituting Eq.~(2.3) and (2.8) to $C_3^{-1} \! \cdot  C_4^{~}$, we obtain
\begin{eqnarray}
P_4^{~} 
&=& 
( U_3^{~} \, \Omega_3^{-1}  A_3^{~T} ) \cdot ( A_4^{~} \, \Omega_4^{~} \, U_4^{~T} )
\nonumber\\
&=& 
U_3^{~} \, \Omega_3^{-1} 
\! \cdot ( A_3^{~T} \! \cdot A_4^{~} ) \, \Omega_4^{~} \, U_4^{~T} 
\nonumber\\
&=&
U_3^{~} \, \Omega_3^{-1} 
\! \cdot {\tilde A}_4^{~} \, \Omega_4^{~} \, U_4^{~T} \, .
\end{eqnarray}
To avoid any confusion, let us write down element of $P_4^{~}$ 
\begin{eqnarray}
&&
P_4^{~}( \sigma' | q_4^{~} | \sigma ) \\
&&
= \sum_{\xi \zeta}^{~}
U_3^{~}( \sigma' | \xi ) \, 
\Omega_3^{-1}( \xi ) \, 
{\tilde A}_4^{~}( \xi \, q_4^{~} | \zeta ) \, 
\Omega_4^{~}( \zeta ) \, 
U_4^{~}( \sigma | \zeta ) \, ,
\nonumber 
\end{eqnarray}
where the new matrix ${\tilde A}_4^{~} = A_3^{~T}  \! \cdot A_4^{~}$ is the 
renormalized orthogonal matrix
\begin{equation}
{\tilde A}_4^{~}( \xi \, q_4^{~} | \zeta ) = \sum_{q_1^{~} \, q_2^{~} \, q_3^{~}}^{~}
A_3^{~}( q_1^{~} \, q_2^{~} \, q_3^{~} | \xi ) \, 
A_4^{~}( q_1^{~} \, q_2^{~} \, q_3^{~} \, q_4^{~} | \zeta ) \, ,
\end{equation}
which satisfies the relation
\begin{equation}
\sum_{\xi q_4}^{~}
{\tilde A}_4^{~}( \xi \, q_4^{~} | \zeta' ) \, 
{\tilde A}_4^{~}( \xi \, q_4^{~} | \zeta ) = \delta( \zeta' | \zeta ) \, .
\end{equation}
In Eq.~(3.4) the group of spins $q_1^{~}$, $q_2^{~}$, and $q_3^{~}$ are
mapped onto the block spin $\xi$ by the RG transformation 
$A_3^{~}$. The obtained ${\tilde A}_4^{~}$ corresponds to the matrix that
constructs MPS, which is constructed by the infinite system DMRG method, 
as shown later.

\begin{figure}
\centerline{\includegraphics[width=5.5cm,clip]{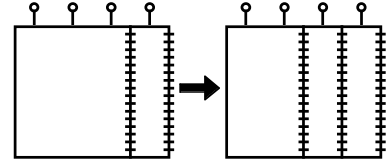}}
\caption{ The area of CTM can be extended by applying the approximate
half-column transfer matrices. }
\label{f4}
\end{figure}

The $P_4$ thus obtained has a function of half-column transfer matrix (HCTM), 
since it extends the width of $C_3^{~}$ by one by way of the dot product
\begin{equation}
C_3^{~} \cdot P_4^{~} = C_3^{~} \cdot 
( C_3^{-1} \! \cdot  C_4^{~} ) = 
( C_3^{~} \, C_3^{-1} ) \cdot C_4^{~} = C_4^{~}
\end{equation}
as shown in the left side of Fig.~4. Applying SVD to $C_3^{~}$ and 
substituting Eq.~(3.2), $C_3^{~} \cdot P_4^{~}$ is calculated as
\begin{equation}
( A_3^{~} \, \Omega_3^{~} \, U_3^{~T} ) \cdot 
( U_3^{~} \, \Omega_3^{-1} \! \cdot {\tilde A}_4^{~} \, \Omega_4^{~} \, U_4^{~T} ) = 
A_3^{~}  \cdot  {\tilde A}_4^{~} \, \Omega_4^{~} \, U_4^{~T} \, .
\end{equation}
Since $C_3^{~}$ is again constructed from $C_2^{~}$ and $P_3^{~}$, 
as shown in the right side of Fig.~4, we can further decompose 
$C_4^{~}$ as
\begin{equation}
C_4^{~} = C_2^{~} \! \cdot \! P_3^{~} \! \cdot \! P_4^{~} =
A_2^{~} \! \cdot \! {\tilde A}_3^{~} \! \cdot \! {\tilde A}_4^{~} \, 
\Omega_4^{~} \, U_4^{~T} \, .
\end{equation}
The contraction process by the dot products are shown in the 
right side of Fig.~5. It should be noted that $C_2^{~} \cdot P_4^{~}$
is not $C_3^{~}$, since $U_2^{~T}$ contained in $C_2^{~}$ and $U_3^{~}$ 
contained in $P_4^{~}$ do not matches to give an identity. In this sense, 
$P_4^{~}$ is an approximation for the half column transfer matrix, 
optimized for the area extension of $C_3^{~}$ only. 

\begin{figure}
\centerline{\includegraphics[width=6.5cm,clip]{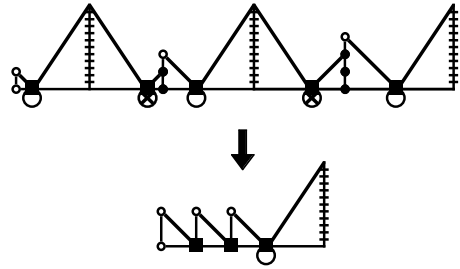}}
\caption{ Pictorial representation of Eq.~(3.8).  }
\label{f5}
\end{figure}

Using the decomposition of $C_4^{~}$ in Eq.~(3.8), we obtain
the matrix product representation of $\Psi_8^{~}  = C_4^{~} C_4^{~T}$. 
We have
\begin{eqnarray}
\Psi_8^{~} 
&=&
A_2^{~} \! \cdot \! {\tilde A}_3^{~} \! \cdot \! {\tilde A}_4^{~} \, 
( \Omega_4^{~} )^2_{~} 
{\tilde A}_4^{~T} \!\! \cdot \! {\tilde A}_3^{~T} \!\! \cdot \! A_2^{~T}
\nonumber\\
&=&
A_2^{~} \! \cdot \! {\tilde A}_3^{~} \! \cdot \! {\tilde A}_4^{~} \, 
\Lambda_4^{~} \, 
{\tilde A}_4^{~T} \!\! \cdot \! {\tilde A}_3^{~T} \!\! \cdot \! A_2^{~T} \, ,
\end{eqnarray}
where $\Lambda_4^{~} = ( \Omega_4^{~} )^2_{~}$ is the 
singular value of $\Psi_8^{~}$. (See Fig.~6.) Such a construction of $\Psi_8^{~}$
is equivalent to the MPS considered in the context of the infinite system
DMRG method. 

\begin{figure}
\centerline{\includegraphics[width=5.5cm,clip]{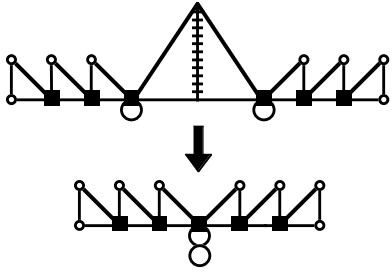}}
\caption{ Matrix product expression of $\Psi_8^{~}$ in Eq.~(3.9). Double
circle represent $\Lambda_4^{~} = ( \Omega_4^{~} )^2_{~}$. }
\label{f6}
\end{figure}

\section{Approximate Area Extension}

\begin{figure}
\centerline{\includegraphics[width=7.5cm,clip]{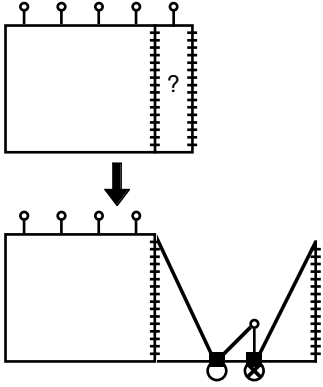}}
\caption{ Approximate area extension process 
$C_5^{App.} = C_4^{~} \cdot {\bar P}_4^{~}$. }
\label{f7}
\end{figure}

Let us consider a problem of obtaining an approximation of
$C_5^{~} = C_4^{~} \cdot P_5^{~}$ without using $P_5^{~}$.
This attempt is equivalent to construct an approximation for 
$C_5^{~}$ using $C_2^{~}$, $C_3^{~}$, or $C_4^{~}$.
One might think that $P_4^{~} = C_3^{-1} \cdot C_4^{~} = 
U_3^{~} \, \Omega_3^{-1} \! \cdot {\tilde A}_4^{~} \, \Omega_4^{~} \, U_4^{~T}$
can be of use as an approximation for $P_5^{~}$. But this idea
should be rejected since $U_4^{~T} U_3^{~}$, which appears in the calculation 
of $C_4^{~} \cdot P_4^{~}$, is not an identity. A way to avoid this
mismatching is to introduce a spatial reflection of $P_4^{~}$, 
which is defined as
\begin{equation}
{\bar P}_4^{~} = 
U_4^{~} \, \Omega_4^{~} \, {\tilde A}_4^{~T} \! \cdot \Omega_3^{-1} U_3^{~T} \, ,
\end{equation}
and use it as an approximation for $P_5^{~}$. 
Leaving the validity of the approximation scheme by the 
latter discussion, let us calculate the approximate extension
$C_5^{App.} = C_4^{~} \cdot {\bar P}_4^{~}$ and write it into
the matrix product representation. (See Fig.~7.) We obtain
\begin{eqnarray}
C_5^{App.} 
&=& 
A_2^{~} \! \cdot \! {\tilde A}_3^{~} \! \cdot \! {\tilde A}_4^{~} \, 
\Omega_4^{~} \, U_4^{~T} \, \, \, 
U_4^{~} \, \Omega_4^{~} \, {\tilde A}_4^{~T} \! \cdot \Omega_3^{-1} U_3^{~T}
\nonumber\\
&=&
A_2^{~} \! \cdot \! {\tilde A}_3^{~} \! \cdot \! {\tilde A}_4^{~} \,
( \Omega_4^{~} )^2_{~}  {\tilde A}_4^{~T} \! \cdot \Omega_3^{-1} U_3^{~T}
\nonumber\\
&=&
A_2^{~} \! \cdot \! {\tilde A}_3^{~} \! \cdot \! {\tilde A}_4^{~} \,
\Lambda_4^{~} \,  {\tilde A}_4^{~T} \! \cdot \Omega_3^{-1} U_3^{~T} \, ,
\end{eqnarray}
and from this approximation we can construct  
\begin{eqnarray}
\Psi_{10}^{App.} &=& C_5^{App.} ( C_5^{App.} )^T_{~} \\
&=&
A_2^{~} \! \cdot \! {\tilde A}_3^{~} \! \cdot \! {\tilde A}_4^{~} \,
\Lambda_4^{~} \, {\tilde A}_4^{~T} \! \cdot \Lambda_3^{-1} \! \cdot \! 
{\tilde A}_4^{~} \, \Lambda_4^{~} \, {\tilde A}_4^{~T} \! \cdot \!
{\tilde A}_3^{~T} \! \cdot \! A_2^{~T} \, , \nonumber
\end{eqnarray}
which may approximate $\Psi_{10}^{~}$. Applying Schmidt 
orthogonalization for $\Lambda_4^{~} {\tilde A}_4^{~T}$ from
the left side
\begin{equation}
\Lambda_4^{~} \, {\tilde A}_4^{~T} = {\tilde B} \, \Lambda
\end{equation}
we obtain the new orthogonal matrix ${\tilde B}$ and the 
right triangular matrix $\Lambda$. (See Fig.~8.) Substituting Eq.~(4.4) into Eq.~(4.3),
we get the matrix product expression
\begin{equation}
\Psi_{10}^{App.} =
A_2^{~} \! \cdot \! {\tilde A}_3^{~} \! \cdot \! {\tilde A}_4^{~} \! \cdot \! 
{\tilde B} \, \Lambda \,  \Lambda_3^{-1} \Lambda_{~}^{\, T} \, 
{\tilde B}_{~}^{\, T} \! \cdot \! {\tilde A}_4^{~T} \! \cdot \! {\tilde A}_3^{~T} 
\! \cdot \! A_2^{~T} \, .
\end{equation}
The extension from $\Psi_8^{~}$ to $\Psi_{10}^{App.}$ is the same as
the wave function extension scheme proposed by McCulloch,~\cite{McCulloch}
where the approximation for the renormalized wave function is given by
\begin{equation}
{\tilde \Psi}_{10}^{App.} =
{\tilde B} \, \Lambda \,  \Lambda_3^{-1} \Lambda_{~}^{\, T} \, {\tilde B}_{~}^{\, T} \, .
\end{equation}
\begin{figure}
\centerline{\includegraphics[width=6.5cm,clip]{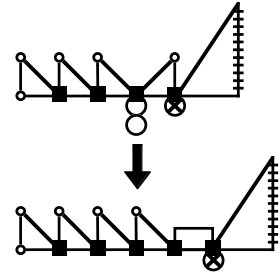}}
\caption{ Reorthogonalization process in Eq.~(4.3). The
rectangular in the lower diagram corresponds to $\Lambda$ 
in Eq.~(4.4).}
\label{f8}
\end{figure}

We have thus obtained a natural explanations for McCulloch's extension scheme from
the view point of 2D vertex model. Up to now we have not considered the
effect of basis truncation, which is used in numerical calculation of the infinite system 
DMRG method. First of all, the extension in Eqs.~(4.4) and (4.5) is still efficient
under the truncation, as it was shown numerically.~\cite{McCulloch}
We then consider the extension from $\Psi_N^{~}$ to ${\tilde \Psi}_{N+2}^{~}$ 
in the large system size limit $N \rightarrow \infty$. For simplicity, let us 
assume that the MPS in this limit is uniform, and the system is away from
criticality. In this limit we can drop the site index from Eq.~(4.1), and can
express the approximate transfer matrix as
\begin{equation}
P =  U \, \Omega_{~}^{-1} \! \cdot {\tilde A} \, \Omega  \, U_{~}^{\, T} 
= U \! \cdot \! {\tilde S} \, U_{~}^{\, T} \, ,
\end{equation}
where ${\tilde S} = \Omega_{~}^{-1} \! \cdot {\tilde A} \, \Omega$. From the 
assumed symmetry of the vertex model, both $P$ and ${\tilde S}$ are symmetric
\begin{eqnarray}
P( \sigma' | \, q \, | \sigma ) &=& P( \sigma | \, q \, | \sigma' ) \nonumber\\
{\tilde S}( \sigma' | \, q \, | \sigma ) &=& {\tilde S}( \sigma | \, q \, | \sigma' ) \, .
\end{eqnarray}
This symmetry is also expressed in short form as ${\bar P} = P$ and 
${\bar {\tilde S}} = {\tilde S}$.
Thus at least when the system size $N = 2i$ is large enough, 
typically several times larger than the correlation length, one can justify 
the usage of $C_i^{~} \cdot {\bar P}_i^{~}$ as the approximation for
$C_i^{~} \cdot P_{i+1}^{~}$.

Before closing this section, we consider the MPS expression for $\Psi_N^{~}$
that is optimized by way of the sweeping process in the finite system DMRG 
method. The matrix product structure
\begin{equation}
\Psi_{N}^{~} 
= 
A_2^{~} \! \cdot \! {\tilde A}_3^{~} \! \cdots \! {\tilde A}_{\frac{N}{2}}^{~} \, 
\Lambda_{\frac{N}{2}}^{~} \, 
{\tilde A}_{\frac{N}{2}}^{~T} \!\! \cdots \! {\tilde A}_3^{~T} \!\! \cdot \! A_2^{~T}
\end{equation}
is similar to that obtained by the infinite system DMRG method, 
but in this case the matrices satisfies the additional relation
\begin{equation}
{\tilde A}_i^{~} \Lambda_i^{~} = \Lambda_{i-1}^{~} {\tilde A}_i^{~T} \, ,
\end{equation}
where both ${\tilde A}_i^{~}$ and $\Lambda_i^{~}$ differ from those
obtained by the infinite system DMRG method. Taking the square root
of $\Lambda_i^{~}$, we formally obtain a diagonal matrix
$\Omega_i^{~} = \sqrt{\Lambda_i^{~}}$. It should be noted that this 
$\Omega_i^{~}$ is different from that obtained from the SVD applied 
to $C_i^{~}$. Defining 
\begin{equation}
{\tilde S}_i^{~} = 
\Omega_{i-1}^{-1} \! \cdot \! {\tilde A}_i^{~} \, \Omega_i^{~} = 
\Omega_{i-1}^{~}  {\tilde A}_i^{~T} \! \cdot  \Omega_i^{-1} 
\end{equation}
%
and substituting it to Eq.~(4.9), we obtain a new standard form for MPS
\begin{equation}
\Psi_{N}^{~} 
= 
\Omega_1^{~} \, 
S_2^{~} \! \cdot \! {\tilde S}_3^{~} \! \cdots \! {\tilde S}_{\frac{N}{2}}^{~} \, 
{\tilde S}_{\frac{N}{2}}^{~T} \!\! \cdots \! {\tilde S}_3^{~T} \!\! \cdot \! S_2^{~T} \, 
\Omega_1^{~} \, ,
\end{equation}
where $\Omega_1^{~}$ is just a constant and is not essential. It is then 
straightforward to obtain the approximation $\Psi_{N+2}^{App.}$ just by
putting  ${\bar {\tilde S}}_{\frac{N}{2}}^{~} \, {\bar {\tilde S}}_{\frac{N}{2}}^{~T}$
at the center of the above MPS, where this insertion is a variant of Eq.~(4.5).
In the thermodynamic limit $N \rightarrow \infty$ the matrix ${\tilde S}_i^{~}$ in Eq.~(4.11) 
is independent on the site index $i$, and therefore it coincides 
with ${\tilde S}$ in Eq.~(4.8). This symmetric representation of uniform 
MPS is often of use.

\begin{figure}
\centerline{\includegraphics[width=4.5cm,clip]{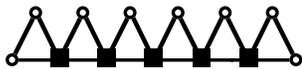}}
\caption{ Graphical representation of Eq.~(4$\cdot$14). }
\label{f9}
\end{figure}

\section{Conclusions and Discussions}

We have considered the wave function prediction in the infinite system 
DMRG method, when it is applied to the 2D vertex model.
Through the singular value decomposition of CTM $C_i^{~}$, we obtained the
approximate half-column transfer matrix $P_i^{~}$. The insertion of 
${\bar P}_i^{~}$ naturally explains the wave function prediction proposed by 
McCulloch,~\cite{McCulloch} which works better than the product wave 
function renormalization group (PWFRG) method,~\cite{PWFRG,PWFRG2,
PWFRG3} especially when the system size is small.
The difference between these two prediction methods can be explained
by the shape of finite size system. The PWFRG method treats growing
triangular cluster,~\cite{PWFRG2} whereas McCulloch's scheme always treat half-infinite
stripe. 

The relation between CTM and MPS in the finite-system DMRG method
is not so clear. For example, $\Psi_8^{~}$ can be expressed as
$C_3^{~} \, C_5^{~T}$, but the MPS representation of the optimized
$\Psi_8^{~}$ by the finite system DMRG cannot be obtained from the
SVD applied to $C_3^{~}$ and $C_5^{~}$ independently. 
This puzzle is something to do with the targeting scheme for 
asymmetric vertex model, and also with the determination of optimal
RG transformation in the real-time DMRG method, where the 
density matrix is time dependent.

\section*{Acknowledgement}

We thank I. McCulloch for valuable comments and discussions.
H.~U. thanks Dr. Okunishi for helpful comments on the DMRG method and 
continuous encouragement. A.~G Acknowledge 
QUTE and VEGA 1/0633/09 grants.

\end{document}